\documentstyle[11pt,epsfig]{article}
\def\lsim{\,\rlap{\lower 3.5 pt \hbox{$\mathchar \sim$}} \raise 1pt
 \hbox {$<$}\,}
\def\gsim{\,\rlap{\lower 3.5 pt \hbox{$\mathchar \sim$}} \raise 1pt
 \hbox {$>$}\,}

\catcode`@=11
\newcount\@tempcntc
\def\@citex[#1]#2{\if@filesw\immediate\write\@auxout{\string\citation{#2}}\fi
  \@tempcnta\z@\@tempcntb\m@ne\def\@citea{}\@cite{\@for\@citeb:=#2\do
    {\@ifundefined
       {b@\@citeb}{\@citeo\@tempcntb\m@ne\@citea\def\@citea{,}{\bf ?}\@warning
       {Citation `\@citeb' on page \thepage \space undefined}}%
    {\setbox\z@\hbox{\global\@tempcntc0\csname b@\@citeb\endcsname\relax}%
     \ifnum\@tempcntc=\z@ \@citeo\@tempcntb\m@ne
       \@citea\def\@citea{,}\hbox{\csname b@\@citeb\endcsname}%
     \else
      \advance\@tempcntb\@ne
      \ifnum\@tempcntb=\@tempcntc
      \else\advance\@tempcntb\m@ne\@citeo
      \@tempcnta\@tempcntc\@tempcntb\@tempcntc\fi\fi}}\@citeo}{#1}}
\def\@citeo{\ifnum\@tempcnta>\@tempcntb\else\@citea\def\@citea{,}%
  \ifnum\@tempcnta=\@tempcntb\the\@tempcnta\else
   {\advance\@tempcnta\@ne\ifnum\@tempcnta=\@tempcntb \else \def\@citea{--}\fi
    \advance\@tempcnta\m@ne\the\@tempcnta\@citea\the\@tempcntb}\fi\fi}
\catcode`@=12

\begin{document}
\vbox to 2 truecm {}
\begin{center}
\boldmath
{\large \bf Large-$p_T$ Inclusive $\pi^{0}$ Cross Sections}  \vskip 3 truemm
{\large \bf  and Next-to-Leading-Order QCD Predictions}   
\vskip 1 truecm
\unboldmath

P. Aurenche,$^1$ M. Fontannaz,$^2$ J.-Ph. Guillet,$^1$ B. A. Kniehl,$^3$
M. Werlen$^1$\\
$^1$ Laboratoire d'Annecy-le-Vieux de Physique Th\'eorique LAPTH,\\
B.P.~110, F-74941 Annecy-le-Vieux Cedex, France\\
$^2$ Laboratoire de Physique Th\'eorique,\footnote{Unit\'e mixte de recherche
(CNRS) UMR 8627} Universit\'e de Paris XI,\\
B\^atiment 210, F-91405 Orsay Cedex, France\\
$^3$ II. Institut f\"ur Theoretische Physik, Universit\"at Hamburg,\\
Luruper Chaussee 149, D-22761 Hamburg, Germany

\end{center}

\vskip 1 truecm
\begin{abstract}

We review the phenomenology of $\pi^0$ production at large transverse
momentum in proton-induced collisions. Uncertainties in the
next-to-leading-order predictions of Quantum Chromodynamics are
discussed. The comparison with data reveals that the disagreement
between theory and experiment lies essentially in an overall
normalization factor.
The situation for $\pi^0$ production is contrasted with that of prompt-photon
production in hadronic collisions.

\end{abstract}

\vfill
\begin{flushright}
hep-ph/9910252\\
LAPTH-751/99\\
LPT-Orsay/99/78\\
DESY 99-153\\
October 1999
\end{flushright}
\newpage
\pagestyle{plain}
\baselineskip=20 pt

\section{Introduction}

The production of hadrons at large transverse
momentum $p_T$ (in hadron-hadron or hadron-photon collisions)
offers a classical test of perturbative QCD. The cross sections of the
hard subprocesses are calculated with next-to-leading-order (NLO)
accuracy \cite {1r}, the quark and gluon distributions in the initial
hadrons are measured in deep-inelastic-scattering experiments
\cite{2r,3r,3rr,4r} and the fragmentation functions describing the transitions
of the partons into the final-state hadrons are extracted
from $e^+e^-$-annihilation data \cite{5r,6r,7r}. Therefore, all the
buil\-ding blocks of the large-$p_T$ cross sections are in
principle known, and the comparison with data provides interesting tests
of the theory.

The hadroproduction of large-$p_T$ $\pi^0$ mesons has been studied by the
CGGRW Collaboration \cite{5r}, who obtained good agreement between the
QCD predictions and collider $\pi^0$ cross sections. On the other hand,
fixed-target data with a lower center-of-mass energy $\sqrt s$
(23~GeV${}\lsim\sqrt s\lsim30$~GeV) overshoot the theoretical predictions,
and it was impossible to obtain an overall agreement with all the
large-$p_T$ $\pi^0$ data. Large-$p_T$ charged-hadron cross sections
were studied in hadroproduction at collider energies \cite{8r} and in
photoproduction at HERA energies \cite{6r}. In that case,
reasonable agreement between theory and data was reached, but one must
remember that these predictions are sensitive to the choice of the
factorization and renormalization scales in the $p_T$-range which
was studied ($p_T\lsim10$~GeV).

In this paper, we come back to the study of the large-$p_T$ $\pi^0$
cross sections. New fixed-target data appeared recently \cite{9r,10r},
completing those already available in the range
$23\lsim\sqrt s\lsim30$~GeV \cite{11r}. On the theoretical side, new
sets of fragmentation function extracted from LEP data are now available
\cite{6r}. One must also notice that $\Lambda^{(4)}_{\overline{\rm MS}}$ has
increased from $\sim$ 200 MeV to $\sim$ 300 MeV since the time of the
first study \cite{5r}. This new value of $\Lambda_{\overline{\rm MS}}$
gives rise to a non-negligible increase of the QCD cross section in
the $p_T$ range studied here ($p_T\lsim10$~GeV).

\begin{figure}[htb]
\begin{center}
\epsfig{file=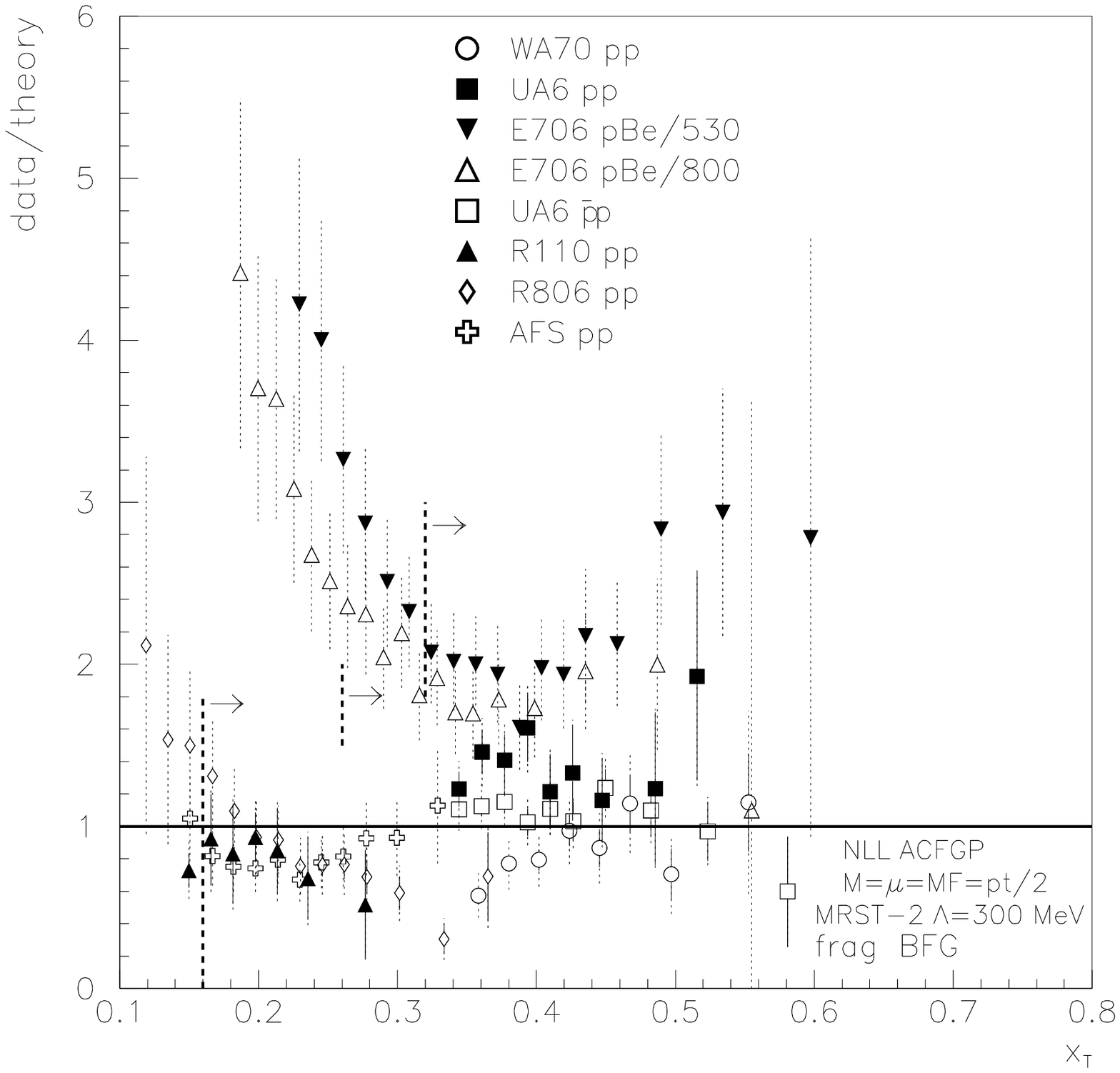,height=12cm,width=16.5cm}
\caption{\em
Dependence on $x_T$ of experimental cross sections for inclusive
prompt-photon production normalized to NLO predictions based on the MRST2
\protect\cite{4r} parton densities.
The dotted vertical lines correspond to $p_T =  5$~GeV/c for the E706 and
ISR experiments.
}
\label{xtcteqpt2}
\end{center}
\end{figure}

Another important reason for this study is the publication of new data
on prompt-photon cross sections \cite{9r,10r}. A recent QCD analysis of
all fixed-target and ISR data \cite{15r} leads to the conclusion that
theory is in reasonable agreement with experiment, with the
exception of two data sets, as shown in Fig.~\ref{xtcteqpt2}.
However, it should be
remembered  that the comparison with QCD predictions can only be done for
high-enough values of $p_T$ ({\it e.g.}\ $p_T\gsim5$~GeV/c for the 
E706 and ISR energy ranges), where the scale
dependence of the theoretical cross sections is not too large. Therefore, in
Fig.~\ref{xtcteqpt2}, one should not consider the first points of the E706
and ISR
experiments. The introduction of an additional intrinsic transverse momentum
$k_T$ of the
incoming partons, strongly advocated in Refs.~\cite{9r,16r} to enforce
agreement between QCD predictions and experiment, is not a very
satisfactory solution; the agreement with the E706 data sets is
improved, but the agreement with other data sets is reduced.

This situation is challenging and leads us to look at
large-$p_T$ $\pi^0$ production. The large-$p_T$ $\pi^0$ mesons form a
significant background for the prompt photons, and their cross section must be
carefully measured in order to allow for a reliable estimate of the
``fake prompt photons,'' in particular due to configurations in which one
photon from
the decaying $\pi^0$ escapes detection. Therefore, there is a strong
experimental correlation between the prompt-photon and
the $\pi^0$ cross sections, especially at low $p_T$ values, where the
$\pi^0$ background is largest. A study of the latter could bring some
clarification on the prompt-photon puzzle.

Section 2 of this paper is devoted to a discussion of the theoretical
calculation of the $\pi^0$ cross sections. We emphasize the uncertainties
associated with the QCD scales, the determination of the fragmentation
functions
and the importance of the higher-order (HO) corrections.
We discuss the comparison
of theoretical predictions with data at fixed-target energies in section 3 and
with data at higher energies in section 4. Section 5 contains the conclusions.

\section{Theoretical Framework}
\label{theory}

At NLO, the inclusive cross section for the
hadroproduction of a single hadron $h$, differential in the transverse
momentum $p_T$ and the pseudorapidity $\eta$ of $h$, takes the following
form:

\begin{eqnarray}
\label{1e}
{d\sigma^h \over d\vec{p}_Td\eta} &=& \sum_{i,j,k=q,g} \int dx_1 dx_2 
F_{i/h_1}(x_1, M) F_{j/h_2}(x_2,M) {dz \over z^2} D_k^h(z,M_F) 
\nonumber \\
&&{}\times \left [\left ( {\alpha_s (\mu ) \over 2 \pi} \right )^2
{d\widehat{\sigma}_{ij,k} \over d \vec{p}_T d\eta}
+ \left ( {\alpha_s(\mu) \over 2 \pi} \right )^3 K_{ij,k}(\mu , M, M_F)
\right ]. 
\end{eqnarray}

The parton densities of the incoming hadrons $h_1$ and $h_2$ are
given by $F_{i/h_1}$ and $F_{j/h_2}$; the fragmentation of a parton $k$
into the hadron $h$ is described by the fragmentation function
$D_k^h(z, M_F)$; $d\widehat{\sigma}_{ij,k}/d\vec{p}_T d\eta$ is the
Born cross section of the subprocess $i + j \to k + X$; and $K_{ij,k}$ is the
corresponding HO correction term.
In this paper, we shall use the ABFOW \cite{2r}, CTEQ
\cite{3r,3rr} and MRST \cite{4r} parton densities. For the fragmentation
functions, we shall use the parametrization of Ref.~\cite{6r}. These
fragmentation functions were derived from fits to charged-pion spectra, but we
make the usual assumption, supported by data, that the rate of $\pi^0$
production is half of that for charged pions.
They are preferable to the parametrizations of Ref.~\cite{5r},
which predate the publication of the LEP hadronic spectra. 

\begin{figure}[htb]
\begin{center}
\epsfig{file=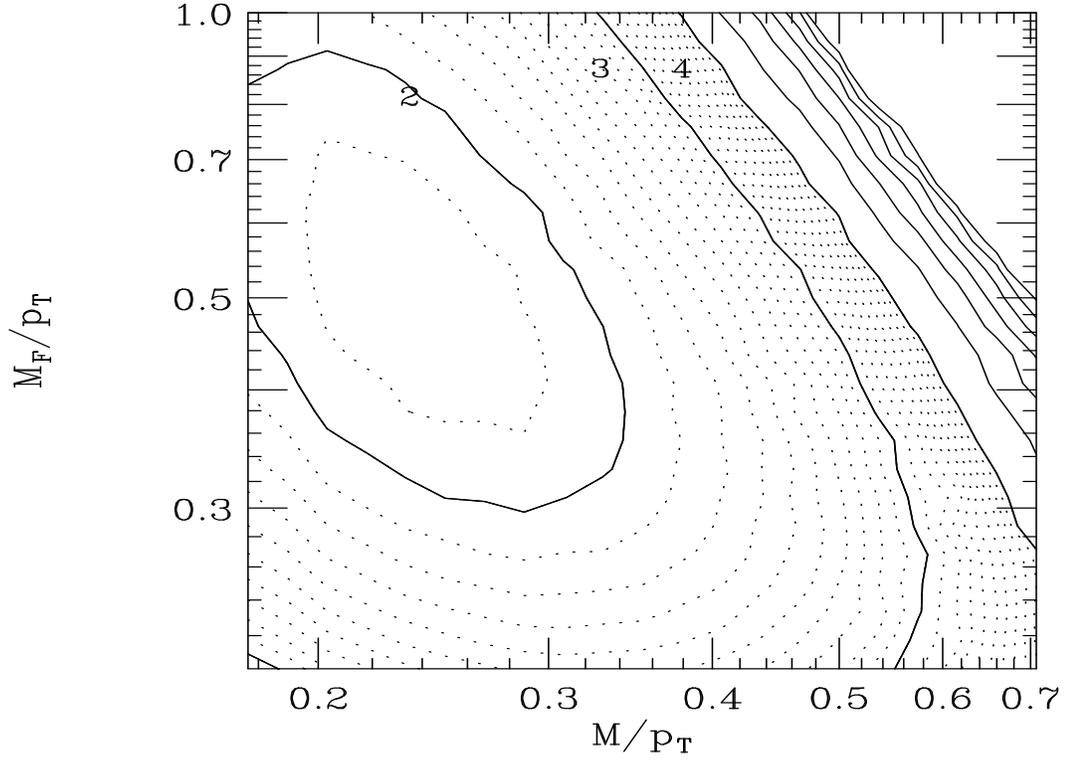,height=10cm,width=14.cm}
\caption{\em
Differential cross section $Ed^3\sigma/d^3p$ for the inclusive production of
$\pi^0$ mesons with $p_T=7$~GeV and $|\eta|<0.75$ in the scattering of
530~GeV protons on a fixed Beryllium target, evaluated as a function of the
initial-state factorization scale $M$ and the fragmentation scale $M_F$
using the MRST2 \protect\cite{4r} parton densities.
The contours of constant cross section are shown in the $(M,M_F)$ plane.
At each point, the renormalization scale $\mu$ is evaluated from
Eq.~(\ref{2e}).}
\label{fig2.ps}
\end{center}
\end{figure}

Expression (\ref{1e}) depends on the initial- and final-state factorization
scales, $M$ and $M_F$, and on the renormalization scale $\mu$. The
cross section $d\sigma^h/d\vec{p}_Td\eta$, calculated to all orders in
$\alpha_s$, is independent of $M$, $M_F$ and $\mu$.  However, the perturbative
series calculated at fixed order in $\alpha_s$ does depend on these scales, the
compensation between the variations of $F_{i/h_1}(M)$, $F_{j/h_2}(M)$,
$D_k^h(M_F)$ and $\alpha_s(\mu)$ and those of $K_{ij,h}(\mu, M, M_F)$ being
incomplete. Therefore, we shall carefully explore the scale
dependence of the cross section (\ref{1e}). On the one hand, we shall use
``standard scales'' $M = M_F = \mu = p_T/\kappa$, with $\kappa$ varying
around $\kappa = 2$, and, on the other hand, we shall fix the scales by using
the Principle of Minimum Sensitivity (PMS) \cite{17r},
\begin{equation}
\label{2e}
\mu\frac{\partial}{\partial\mu}\,\frac{d\sigma^h}{d\vec{p}_Td\eta}=
M\frac{\partial}{\partial M}\,\frac{d\sigma^h}{d\vec{p}_Td\eta}=
M_F\frac{\partial}{\partial M_F}\,\frac{d\sigma^h}{d\vec{p}_Td\eta}=0,
\end{equation}
which determines the optimal scales $\mu_{\rm opt}$, $M_{\rm opt}$ and
$M_{F,\rm opt}$.
An illustration of the scale sensitivity is presented in Fig.~\ref{fig2.ps},
where the differential cross section for the inclusive production of single
$\pi^0$ mesons, with $p_T= 7$~GeV/c and $|\eta|<0.75$, through the
scattering of 530~GeV/c protons on a fixed Beryllium target is considered.
Specifically, the contours of constant cross section are shown in the
$(M,M_F)$ plane.
For each set $(M,M_F)$, $\mu$ is determined by the first equality in
Eq.~(\ref{2e}).
As expected, one finds a minimum, corresponding to the optimal
choice defined by Eq.~(\ref{2e}), but one also notices that the area
of stability under changes of scales around this optimal point is not
very large. It is smaller than in the case of prompt-photon production.
We noticed in several numerical studies that it is not possible
to verify the PMS criterion (\ref{2e}) for too small values of $p_T$,
typically for $p_T\lsim5$~GeV/c. This indicates that the HO corrections are 
large and that the QCD predictions are quite unstable under changes of scales.
Therefore, we shall emphasize the comparison between theory and experiment at
$p_T\gsim5$~GeV/c, although we shall also explore smaller $p_T$ values.

Since working in the optimal scheme is rather cumbersome and requires a lot of
numerical work, we compare, in general, theory predictions for fixed scales
with experiment. We use the standard choices $M = M_F = \mu = p_T/ 2$ and
$p_T/ 3$.
As will be discussed later, the optimal scales, derived from Eq.~(\ref{2e}),
turn out to interpolate between these two choices, depending on the
kinematical range considered.

Besides the scale dependence, there are other sources of uncertainties
in Eq.~(\ref{1e}), which come from the parton densities and the
fragmentation functions. The parton densities are generally determined
with good accuracy. However, at
small $p_T$ and for scale $p_T/3$, 
the factorization scale may approach the starting scale $Q_0$ of the 
QCD evolution. In this region, the parton densities are not
constrained by data. To avoid this problem when using the scale
$p_T/3$, we thus require $p_T$ to be larger than 4.5~GeV/c.

Another source of uncertainty concerns the fragmentation functions. The quark
fragmentation is now well constrained by PETRA, PEP and LEP data \cite{6r,7r}.
However, the dominant support to Eq.~(\ref{1e}) comes from the large-$z$
domain, where the $e^+ e^-$ data are scarce. For instance, numerical studies
of Eq.~(\ref{1e}) indicate that the mean $z$ value is about 0.85 for
$4<p_T<8$~GeV/c and $\sqrt s = 31$~GeV. There are very few data
points in the
range $0.8<z<1$ (zero or one point in most data sets), and they have large
error bars. Therefore, we expect the uncertainty coming from the fragmentation
functions to be of the order of a few tens of percents.
It must also be noticed that the gluon fragmentation function is not well
determined by $e^+e^-$ data, since it appears there only at NLO.
More constraints could, in principle, be obtained from inclusive pion
production at large transverse momentum in hadronic collisions.
For example, the inclusive large-$p_T$ charged-hadron cross section has
been measured by the UA1 Collaboration \cite{18r} in the range
$5\lsim p_T\lsim20$~GeV/c. These data are compatible with the
parametrization of Ref.~\cite{6r}, but they are also compatible with the same
parametrization after the normalization of the gluon fragmentation function
has been increased by 30\% to 40\%. (Such a large change in the gluon
fragmentation function is still compatible with the $e^+e^-$ data.)
This gives us an estimate
of the flexibility in the gluon parametrization. We remark that changing the
gluon fragmentation function may affect the slope of the $p_T$
distributions,
since the low $x_T$ regions will be more strongly affected. Besides these
problems
related to the $z$ behavior of the fragmentation functions, we should remark
that we are led to explore small fragmentation scales (in particular, when
using $M_F = p_T/3$) far away from the kinematical regions where the fits
are performed. In summary, hadronic data require large $z$ values and small
scales, while the fragmentation functions are (mostly) extracted from
$e^+e^-$ data at medium $z$ values and large scales.

Finally, let us mention that we do not use theoretical expression in
which the large $\ln(1 - z)$ terms present in the HO corrections are
resummed \cite{19r,20r}. Indeed, in order to be coherent, one should also
use resummed expressions when extracting fragmentation functions from
$e^+e^-$ data. Such an analysis has not been done so far. It would
certainly be very interesting to pursue phenomenological studies in this
direction \cite{21r}. One must, however, keep in mind that the
optimization procedure of Stevenson and Politzer \cite{17r} amounts to a
partial resummation of the $\ln(1 - z)$ terms \cite{22r}.

For all these reasons, we do not expect a very good agreement between
fixed-target data and QCD predictions. The expected agreement should
only be within a few tens of percent. At larger energies, the mean value
of $z$ is smaller and the sensitivity to the scale variations is reduced. A
better agreement is therefore expected and has been verified in previous
studies \cite{5r,8r}. This point will be discussed later.

In this paper, all the calculations are performed in the $\overline{\rm MS}$
scheme. The value of $\Lambda_{\overline{\rm MS}}$ in $\alpha_s$, calculated
at two loops, is taken to be equal to the one used in the parton densities.

\section{Fixed-Target Data and Theory}

\boldmath
\subsection{Inclusive Distributions in $p_T$}
\unboldmath

\begin{figure}[htb]
\begin{center}
\epsfig{file=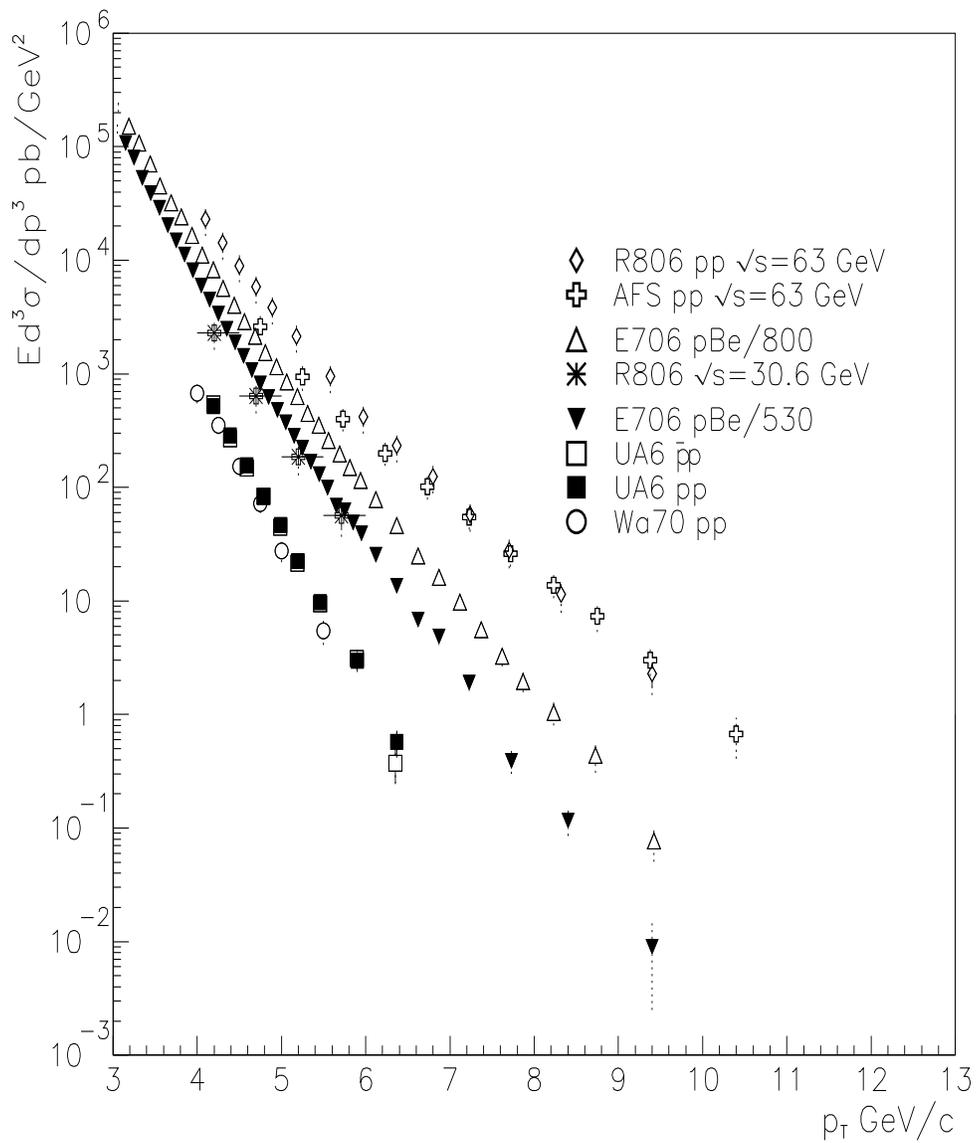,height=17.cm,width=14.cm}
\caption{\em
Compilation of the inclusive $\pi^0$ cross sections discussed in the text
as functions of $p_T$.
}
\label{allpio}
\end{center}
\end{figure}

The data that we are going to discuss are displayed in Fig.~\ref{allpio}.
Unlike what was observed in the case
of prompt-photon production, the $\pi^0$ data taken at the same energy
are quite compatible with each other, with the exception of ISR
data at $\sqrt s = 63$~GeV, where the R806 \cite{12r8} and AFS \cite{13r} 
data disagree
for $p_T\lsim6$~GeV/c.
The increase in cross sections as $\sqrt s$ increases at fixed
$p_T$ is also visible. Let us emphasize that we only consider $pp$,
$p\bar{p}$ and $pBe$ data in order to reduce the uncertainties due to
the parton densities of the incoming hadrons. We do not study $\pi p$
data, but concentrate on proton-induced fixed-target data and ISR
data in the energy range below $\sqrt s = 63$~GeV.

In what follows, we consider only the range $p_T > 4$~GeV/c, which
is covered by almost all data sets. With this cut, we also efficiently
suppress possible non-perturbative contributions to the cross sections,
such as power corrections, intrinsic $k_T$ effects, {\it etc}.

\begin{figure}[htb]
\begin{center}
\epsfig{file=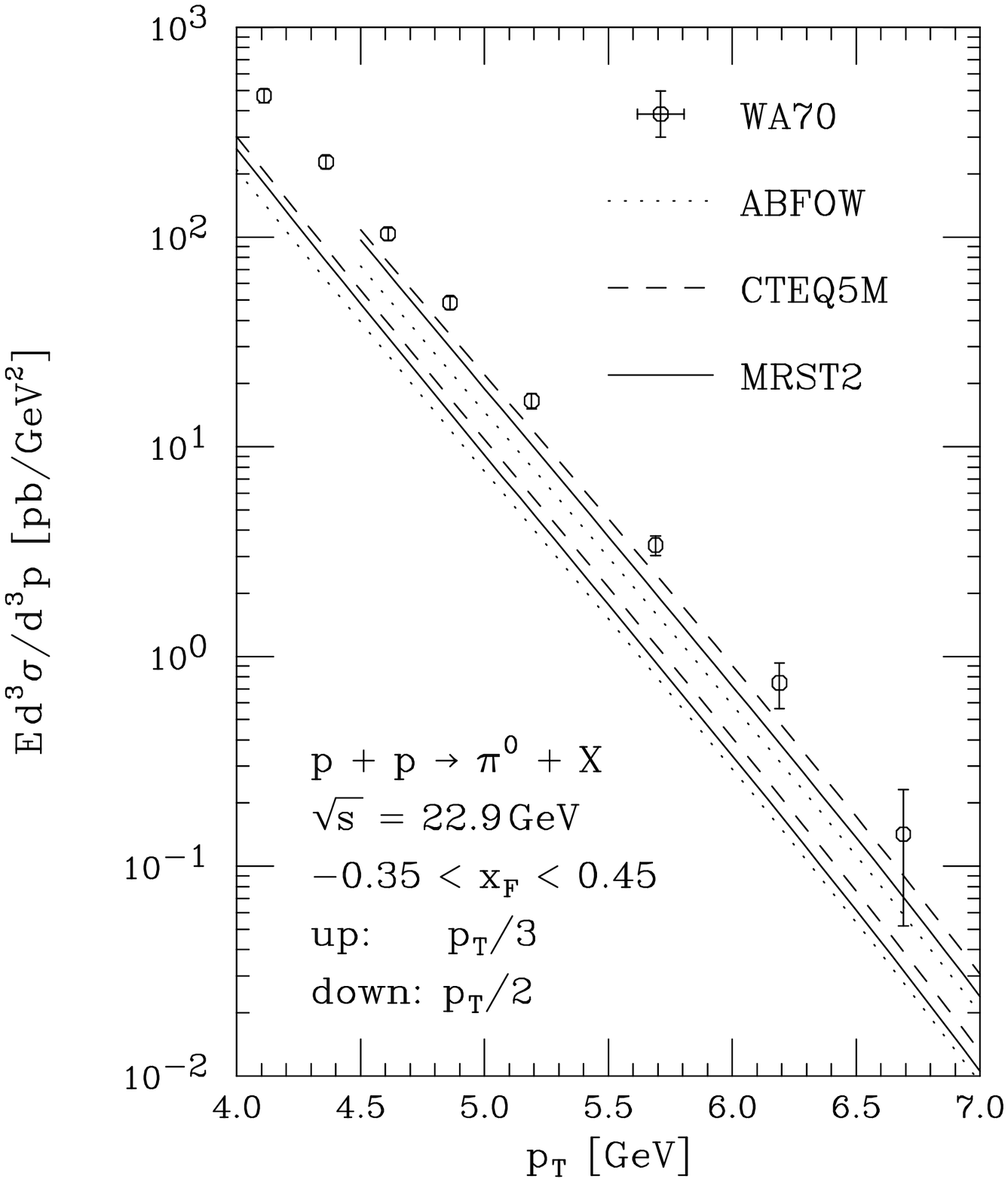,height=9.cm,width=9.cm}
\caption{\em
Comparison of WA70 \protect\cite{11r} $\pi^0$ data with NLO predictions for
three different sets of parton densities and two different scale choices.
The statistical and systematic errors are added in quadrature.
}
\label{fig4ps}
\end{center}
\vskip -1.5cm
\begin{center}
\epsfig{file=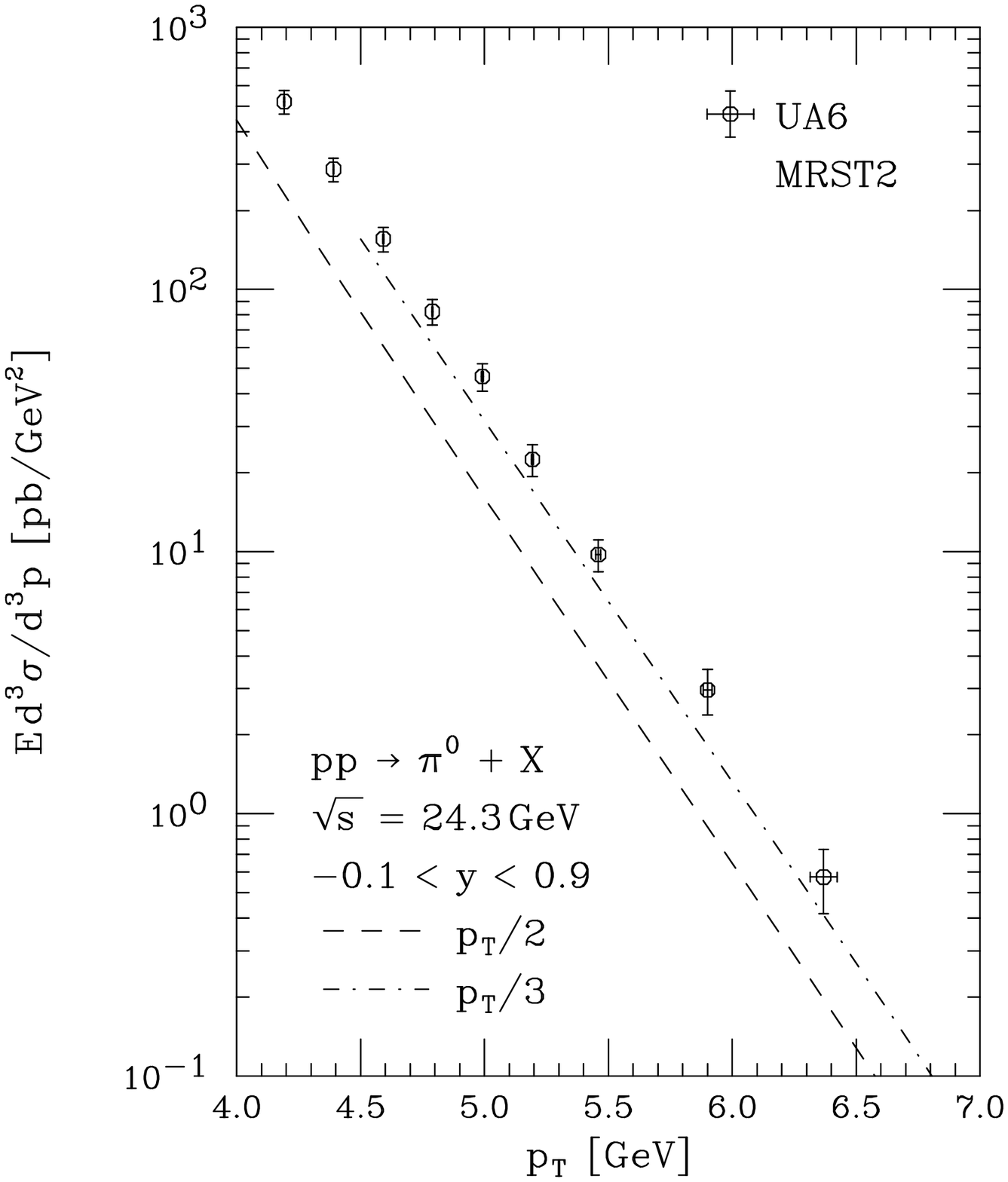,height=9.cm,width=9.cm}
\caption{\em
Comparison of UA6 \protect\cite{10r} $\pi^0$ data with NLO predictions based
on the MRST2 \protect\cite{4r} parton densities.
Very similar results are obtained with the CTEQ4M \protect\cite{3r} parton
densities.
}
\label{fig5ps}
\end{center}
\end{figure}

In Fig.~\ref{fig4ps}, we show the effects of the variation in the parton
densities and scales on the predictions for the WA70 experiment \cite{11r}.
For both scale choices, $p_T/2$ and $p_T/3$, we observe the
following hierarchy:
ABFOW and CTEQ5M lead to the smallest and largest predictions, respectively,
while MRST2 leads to an intermediate result.
The result for CTEQ4M almost coincides with that for MRST2 and is not shown
in Fig.~\ref{fig4ps}.
This hierarchy may be explained by the different $\Lambda_{\overline{\rm MS}}$
values used in the evaluations of $\alpha_s$.
For four quark flavors, these are 230 MeV for ABFOW, 296 MeV for CTEQ4M,
300 MeV for MRST2 and 326~MeV for CTEQ5M.
When $\alpha_s$ is evaluated with a common value of
$\Lambda_{\overline{\rm MS}}$, the four results become very similar.
The effect of the scale choice is also clearly displayed. One notices a
$p_T$-independent increase by roughly a factor of two when the scale is
reduced from $p_T/2$ to $p_T/3$.
This is a typical example for the sensitivity of the theory to the scale
choice. As we shall see later, the results obtained 
using optimal scales are close to those obtained with scale $p_T/3$.
As expected, the agreement between data and theory is not very good. With
small scales ($\sim p_T/3$), the theory underestimates the data by some
40\% to 50\%.
In Fig.~\ref{fig5ps}, UA6 $pp$ data \cite{10r} are compared with the 
corresponding QCD predictions. 
For this experiment, the QCD predictions for scale $p_T/3$ undershoot the
data by some 30\%.

\begin{figure}[htb]
\begin{center}
\epsfig{file=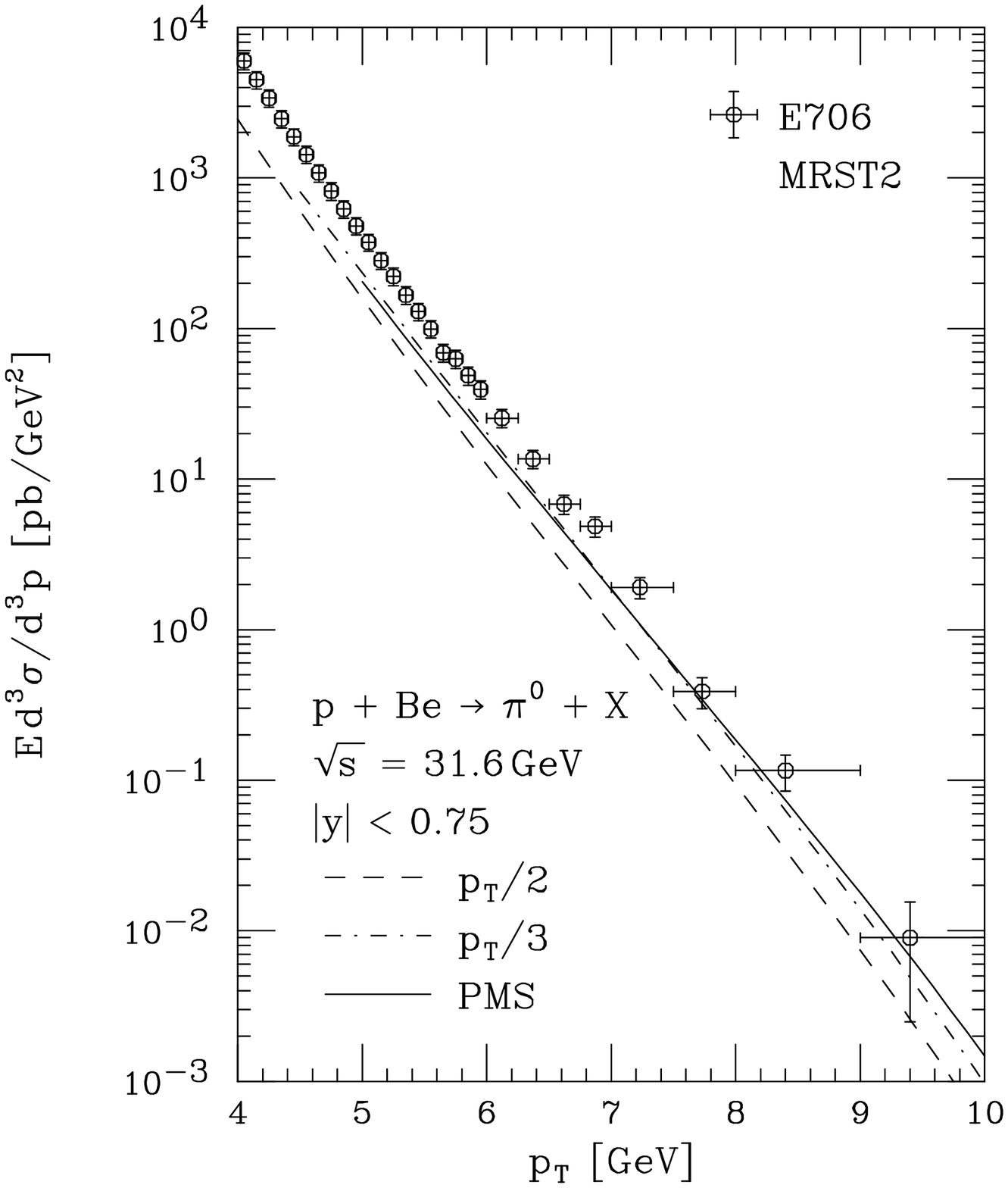,height=9.cm,width=9.cm}
\caption{\em
Comparison of E706 \protect\cite{9r} $\pi^0$ data at $E=530$~GeV
with NLO predictions based on the MRST2 \protect\cite{4r} parton densities.
}
\label{fig6ps}
\end{center}
\vskip -1.5cm
\begin{center}
\epsfig{file=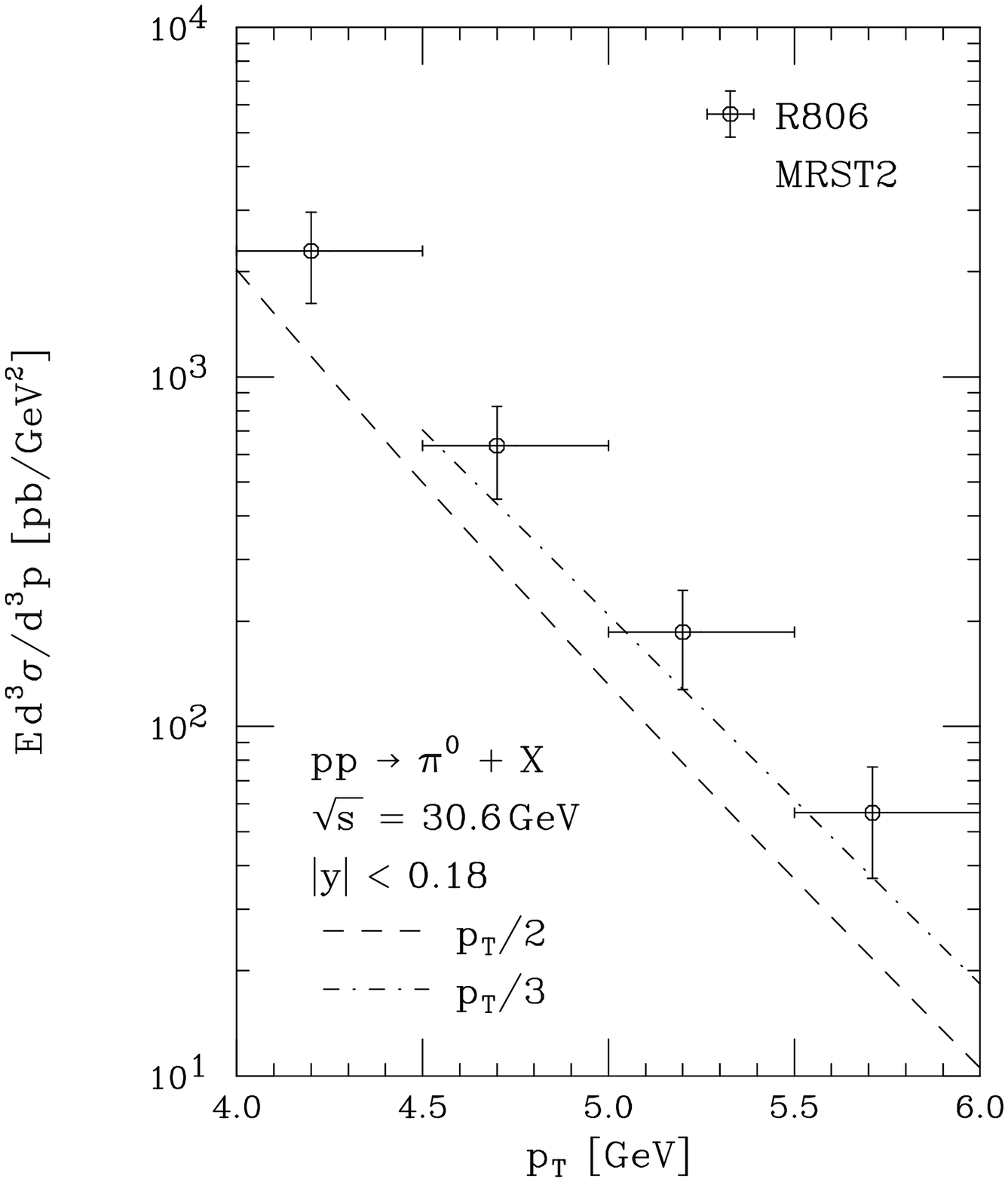,height=9cm,width=9cm}
\caption{\em
Same as Fig.~\protect\ref{fig5ps}, but for R806 \protect\cite{12r8} $\pi^0$
data at $\protect\sqrt s=30.6$~GeV.
}
\label{fig7ps}
\end{center}
\end{figure}

Similar remarks hold for the E706 data \cite{9r}.
In Fig.~\ref{fig6ps}, we display data and predictions for $\sqrt s =
31.6$~GeV. One observes that the curve corresponding to the optimal
scales (labelled PMS) is very close to the one obtained with scale $p_T/3$.
Here again, the disagreement between theory and data, for scale $p_T/3$
or optimal scales, is of the order of a few tens of percents. It is
interesting to notice that, with scale $p_T/2$, which allows one to
extend the predictions down to smaller values of $p_T$, the theoretical
cross section is almost parallel to the data.
Therefore, there is no such increase
of the ratio of data over theory as was observed in the prompt-photon case (see
Fig.~\ref{xtcteqpt2}). Very similar conclusions can be drawn from the 
comparison of the E706 data at $\sqrt s =38.8$~GeV with theory.

Finally, in Fig.~\ref{fig7ps}, we show the ISR R806 data at $\sqrt s=30.6$~GeV
\cite{12r8} and the theoretical predictions obtained with scales $p_T/2$
and $p_T/3$. Here again, small scales lead to the best agreement with the
data.

From this short survey of data and QCD predictions we can draw some preliminary
conclusions. As expected, we do not obtain a very good agreement
with respect to normalization. The predictions for scale $p_T/3$
systematically underestimate the data by some 30\% to 50\% for all
considered experiments. On the other hand, the $p_T$ behavior is well
reproduced. One also notices that all data appear to be consistent with each
other. 

\boldmath
\subsection{Inclusive Distributions in $x_T$}
\unboldmath

\begin{figure}[tb]
\begin{center}
\epsfig{file=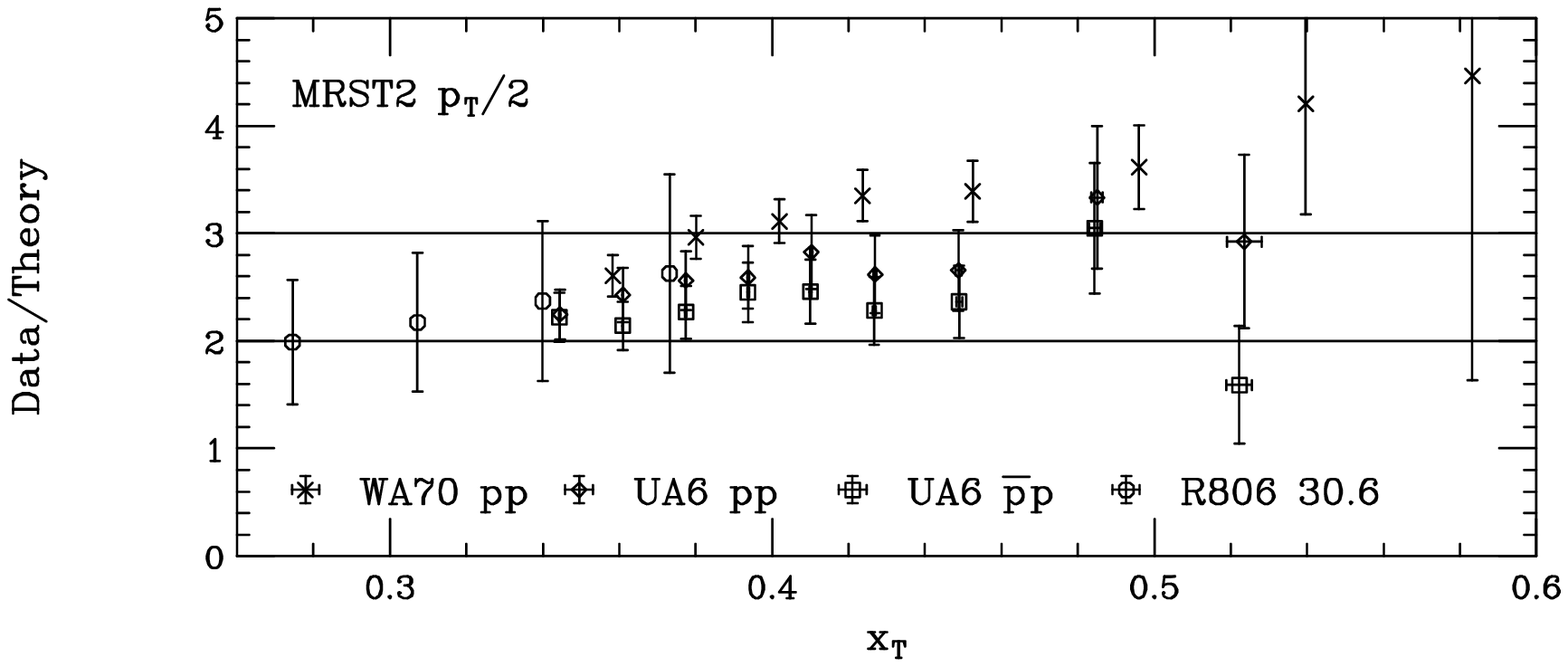,height=7.5cm,width=15.cm}
\caption{\em
Comparison of WA70 \protect\cite{11r}, UA6 \protect\cite{10r} and ISR
\protect\cite{12r8} $\pi^0$ data with NLO predictions based on the MRST2
\protect\cite{4r} parton densities.
All scales are set equal to $p_T/2$ with $p_T=2x_T\protect\sqrt s$.
The horizontal lines drawn at 2 and 3 illustrate the mutual agreement of all
data sets to within $\pm20\%$.}
\label{fig8ps}
\end{center}
\vskip-1.cm
\begin{center}
\epsfig{file=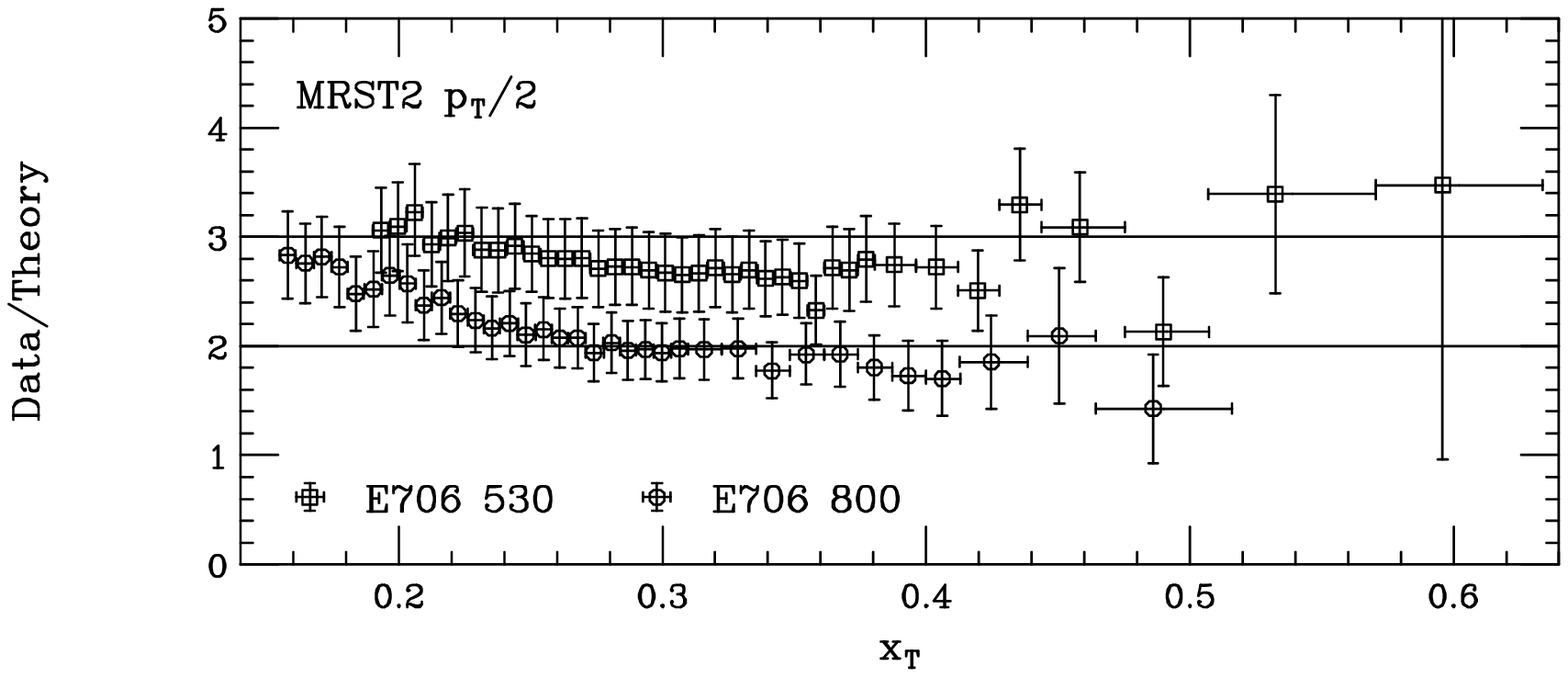,height=7.5cm,width=15.cm}
\caption{\em
Same as Fig.~\protect\ref{fig8ps}, but for E706 \protect\cite{9r} $\pi^0$
data.
}
\label{fig9ps}
\end{center}
\end{figure}

The  above discussion is best summarized by displaying on a linear scale the
data as differential cross sections in $x_T$ normalized to the theoretical
predictions. The data, normalized to the NLO predictions evaluated using set
MRST2 and a common scale set equal to $p_T/2$, are displayed in
Figs.~\ref{fig8ps} and \ref{fig9ps}.
We notice that the normalized data are mutually compatible within $\pm20\%$,
and that multiplying the theoretical predictions by a common normalization
factor of 2.5 will bring data and theory in reasonable agreement. 
It is also interesting to remark that the ratio of data over theory is rather
flat and shows no such sharp rise at low $x_T$ as was observed in the
prompt-photon production experiments (see Fig.~\ref{xtcteqpt2}).

\begin{figure}[htb]
\begin{center}
\epsfig{file=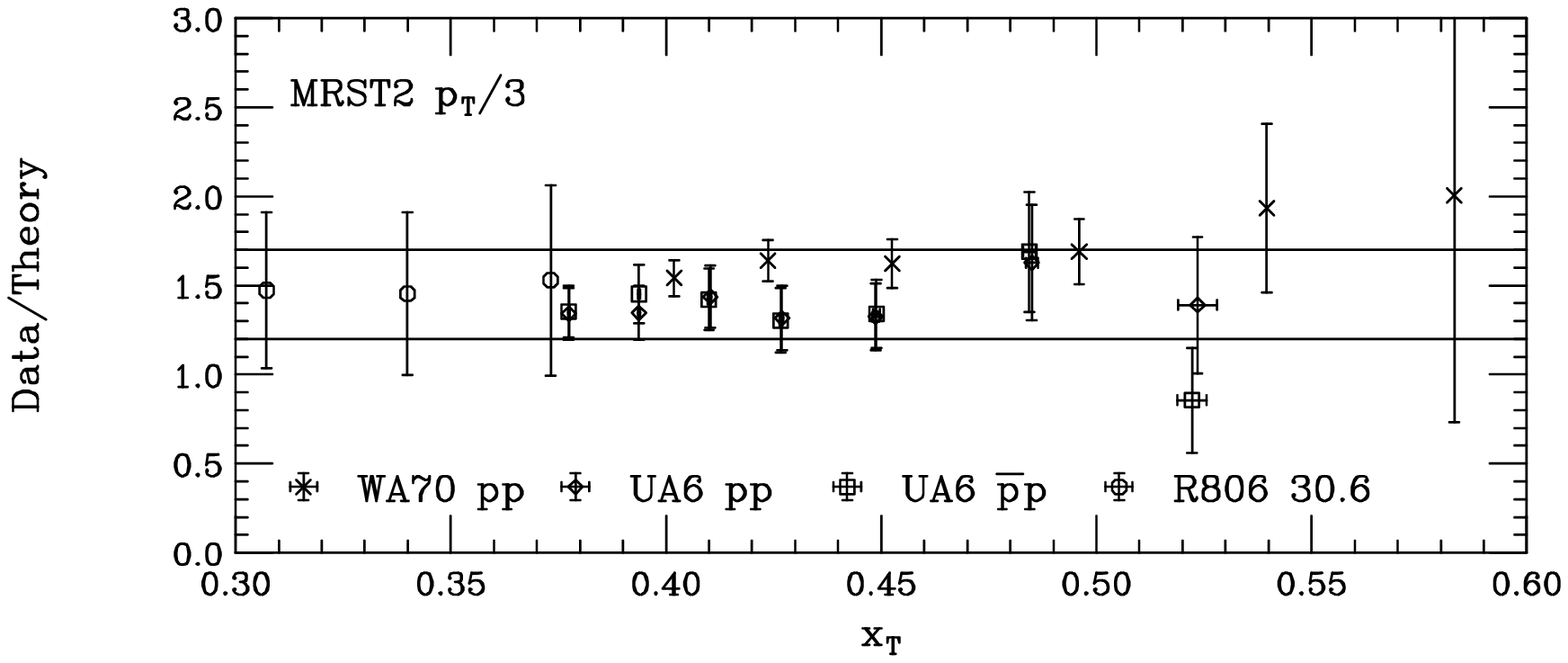,height=7.5cm,width=15.cm}
\caption{\em
Same as Fig.~\protect\ref{fig8ps}, but for scale choice $p_T/3$.
The horizontal lines drawn at 1.2 and 1.7 illustrate the mutual agreement of
all data sets to within $\pm20\%$.
Only data points with $p_T>4.5$~GeV are kept to avoid the use of too small
factorization scales in the NLO predictions.
}
\label{fig10ps}
\end{center}
\vskip-1.cm
\begin{center}
\epsfig{file=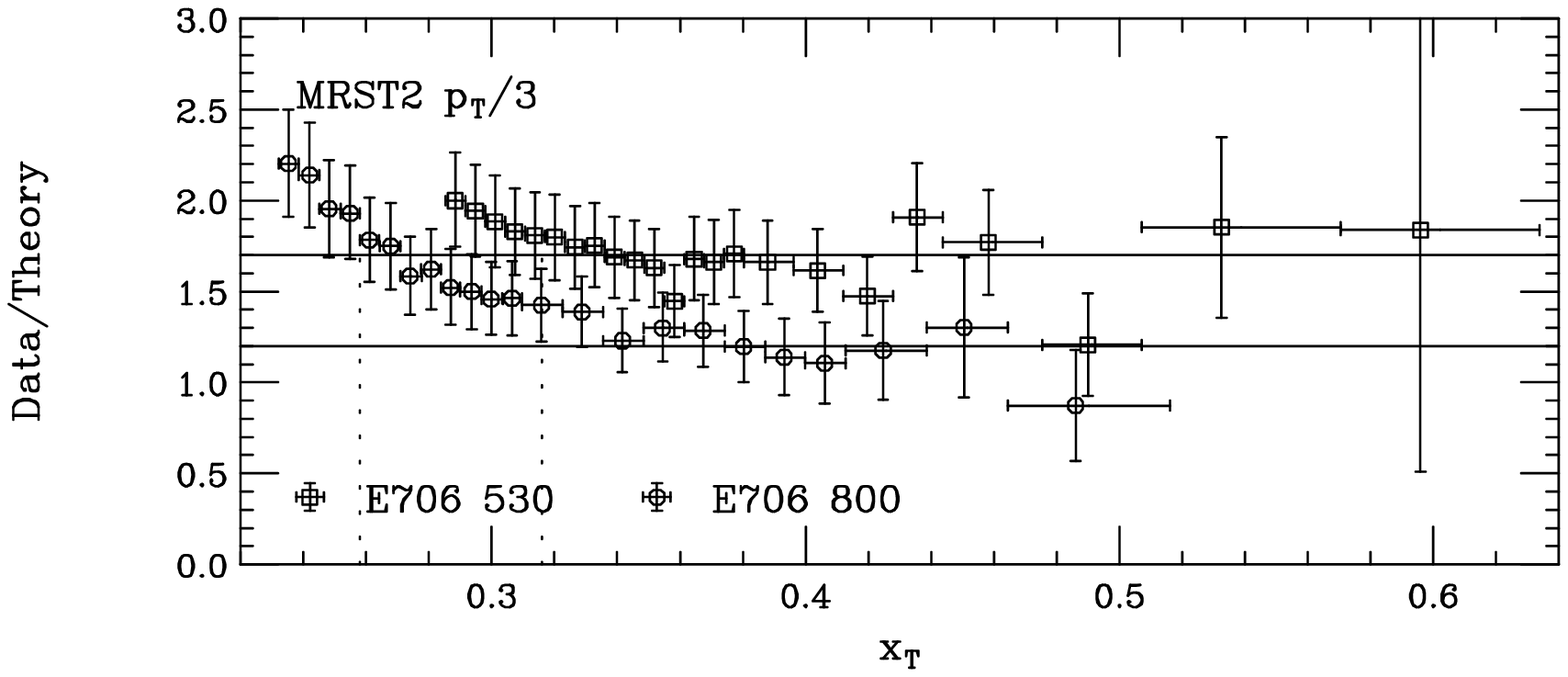,height=7.5cm,width=15.cm}
\caption{\em
Same as Fig.~\protect\ref{fig10ps}, but for E706 \protect\cite{9r} $\pi^0$
data.
The dotted vertical lines correspond to $p_T=5$~GeV/c for the two E706
energies.
Only data points with $p_T>4.5$~GeV are kept to avoid the use of too small
factorization scales in the NLO predictions.
}
\label{fig11ps}
\end{center}
\end{figure}

These general features are also found when comparing data and theory with the
common scale put equal to $p_T/3$.
The results are shown in Figs.~\ref{fig10ps} and \ref{fig11ps} for the
MRST2 parton distributions. The major
difference between theory and experiment now resides in a normalization
factor of about 1.45. Comparing Fig.~\ref{fig11ps} with
Fig.~\ref{fig9ps}, one notices the different slope of the data-over-theory
ratio at small $x_T$ values, corresponding to $p_T \lsim 5$~GeV/c,
indicating that the theoretical predictions are rather unstable in this region
and cannot be trusted.

\subsection{Discussion}

Based on the above analysis of fixed-target data (WA70, UA6, E706, R806 at 
$\sqrt s =30.6$~GeV), we reach the following conclusions concerning inclusive
single-$\pi^0$ production. All data appear to be consistent with each other 
to within $\pm20\%$. However, they differ from the fixed-scale theoretical 
predictions by a rather large normalization factor, namely $K=2.5 \pm 0.5$ for
scale $p_T/2$ and $K=1.45 \pm 0.25$ for scale $p_T/3$.
In the latter case, it is worth investigating
if this  normalization factor could be accounted for, in the theoretical
calculations, by a different choice of fragmentation functions, which are
still rather flexible in the kinematical range of interest, as explained in
section~\ref{theory}.

This situation is to be compared to that of prompt-photon production, where
the scattering of data sets is much larger. In fact, normalized to the
theoretical predictions for fixed scales, the measured E706 rate at
$\sqrt s =31.6$~GeV appears two to four times larger than the corresponding
WA70 rate.

Another feature distinguishes the $\pi^0$ spectra from the prompt-photon
spectra when they are compared to theory.
For scale $p_T/2$, the data-over-theory ratio for the $\pi^0$ data
is rather flat down to rather low $x_T$ values (see
Fig.~\ref{fig9ps}), while the corresponding ratio for the prompt-photon data
is found to exhibit a sharp rise for decreasing $x_T$.
For scale $p_T/3$, a rise is observed for both $\pi^0$ and prompt-photon
data, the rise being sharper in the latter case.
This change of behavior signals an instability of the theory: for $\pi^0$
production at E706 energies, a change of scales introduces a change of slope in
the $p_T$ spectrum. For WA70 or UA6 energies, the change of scales reduces
to a change in the overall normalization.

A final remark concerns the energy variation of the $\pi^0$ spectrum. By
comparing data at different energies in the very same experiment, we are able
to better examine the variation with $\sqrt s$ of the theoretical
predictions. This
can be done by looking at Figs.~\ref{fig9ps} and \ref{fig11ps}, where E706
data at $\sqrt s = 31.6$~GeV and 38.8~GeV are displayed. We
notice a
30\% to 40\% difference in normalization, which could indicate that the
theoretical predictions calculated with fixed scales are not able to follow
the energy dependence of the data. The use of optimized scales does not
improve the agreement with data, since the predictions based on scale
$p_T/3$ or optimized scales give very similar results for all data with 
energy below $\sqrt s\approx40$~GeV.

\section{Higher-Energy Data and Theory}

\begin{figure}[htb]
\begin{center}
\epsfig{file=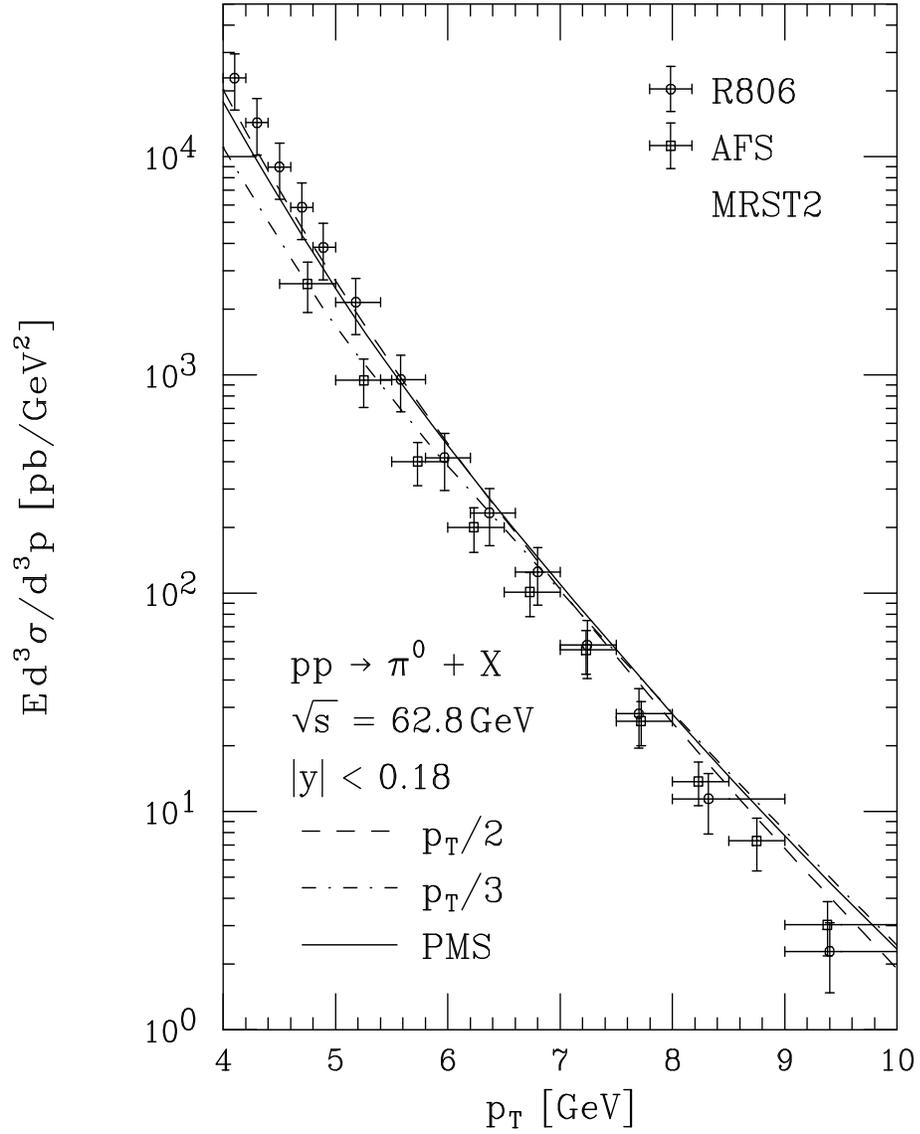,height=15.cm,width=12.cm}
\caption{\em
Comparison of R806 \protect\cite{12r8} and AFS \protect\cite{13r} $\pi^0$ data
with NLO predictions based on the MRST2 \protect\cite{4r} parton densities.
}
\label{fig13ps}
\end{center}
\end{figure}

In Fig.~\ref{fig13ps}, we compare two sets of ISR data at 63~GeV, from the
R806~\cite{12r8} and AFS~\cite{13r} Collaborations, with theoretical
predictions.
In contrast to the results for lower energies, we observe a remarkable
stability in the theoretical predictions for $p_T\gsim5$~GeV/c
($x_T\gsim0.16$): all our scale choices lead to similar predictions.
Below $p_T\approx5$~GeV/c, the predictions diverge.
The choice $p_T/2$ and the use of optimized scales give similar results, while
the choice $p_T/3$ leads to a cross sections which is smaller by a factor of
two.
Below $p_T\approx6$~GeV/c, no conclusion is possible because each data set
favours a different scale choice. Given the statistical relevance of the
R806 data, the choice $p_T/2$ or the use of optimized scales appear to be
most appropriate.

The agreement of theory with both data sets is rather good for
$p_T\gsim 6$~GeV/c, although the theoretical expectations tend to be
somewhat higher than
the data, a situation which is different from the fixed-target case. This fact
may be related to the observation made above concerning the energy variation
of the data-over-theory ratio. This can be checked, to an extent which is 
limited by the relatively large error bars, by comparing the fixed-scale
theory to the three sets of R806 data at $\sqrt{ s} = 30.6$, 44.8 and
62.8~GeV. We find that, at
fixed $x_T$, this ratio tends to decrease when the energy increases, in
agreement with the E706 results.

Turning to the UA1 data on charged-pion production \cite{18r}, similar
comments as above can be made.
The theoretical predictions at large $p_T$ values, above $p_T\approx7$~GeV/c,
are relatively stable and agree well with the data, while, at lower $p_T$
values, the predictions start to diverge, thus bracketing the experimental
points.

\section{Conclusions}

The fairest conclusion to be drawn from our studies of pion production in
hadronic collisions is that the phenomenology of this process is not yet
completely understood! Several problems can be identified.

On the theoretical side, the main difficulty lies in the scale instability,
which is significant at low energy but disappears for $\sqrt s\gsim60$~GeV, at
least for large-enough $p_T$ values. Hopefully, this instability will be
partly removed by resumming the large $\ln(1-z)$ terms associated with the
fragmentation process, very much in the way the resummation of threshold
factors improved the predictions of prompt-photon production at large
transverse momentum \cite{19r,20r,21r}. This improvement would be important in
order to
better understand the energy dependence of the cross sections, which, as we
have seen, cannot be fully understood when using the fixed-scale approach.
In the meantime, the choice of optimized scales, which is equivalent to 
setting the scales equal to $p_T/3$ at fixed-target energies and to
$p_T/2$ at upper ISR energies, seems
to be the most appropriate one and is anticipated to give results not too far
from the resummed theory ({\em cf.}\ the case of prompt-photon production).

On the phenomenological side, one needs a second generation of fragmentation
functions, based on the resummed approach and also taking account of pion
spectra in (selected) hadronic collisions. From the $e^+ e^-$ data, upon which
existing parametrizations are based, the gluon to pion fragmentation is
largely undetermined, and there are very few experimental points to directly
constrain the quark to pion fragmentation at very large $z$ values.
There is probably enough flexibility in the fragmentation functions to change
the size of the predictions for hadronic collisions by 30\% or so.
This point is certainly worth a more detailed investigation.

Comparing theory with data, the fixed-target data are found to lie
systematically above the theory predictions, while the ISR data are compatible
with or somewhat below the predictions and the UA1 data are in perfect
agreement with the predictions. This problem of energy dependence of
the cross section at fixed $x_T$ is particularly manifest in the case of
the E706 data. Is this an experimental problem? Is the scale dependence of the
theory more involved than expected? Redoing the phenomenology using the
resummed approach will probably help to better understand this point.

Although confused, the case of $\pi^0$ production is still much simpler than
that of prompt-photon production, where 24~GeV and 63~GeV data are rather
compatible with the predictions, while 30~GeV and 40~GeV data are much larger.
Considering only experiments below 40~GeV and taking into account the fact
that the $\pi^0$ data taken in those experiments are compatible with each 
other, this seems to indicate that
the systematic errors on prompt-photon production are probably underestimated.
In particular, the background subtraction necessary to obtain the
prompt-photon spectrum should be carefully re-assessed.

It is important for the search of the Higgs-boson decay to two photons to
understand the production of photons and pions at LHC energies. Given the
confused situation at lower energies, after more than twenty years of intense
experimental and theoretical efforts, a lot of work remains to be done to
achieve this goal.

\newpage

\begin{figure}[htb]
\begin{center}
\epsfig{file=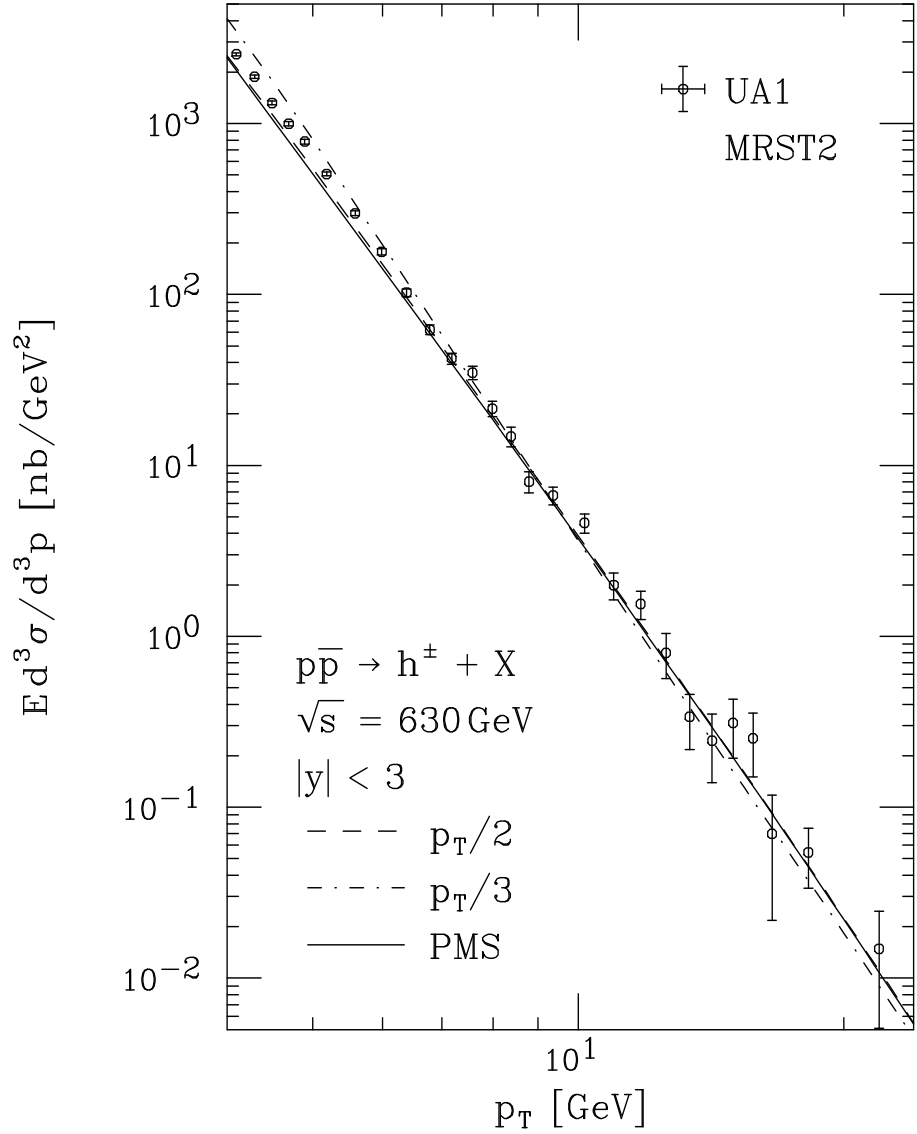,height=15.cm,width=12.cm}
\caption{\em
Comparison of UA1 \protect\cite{18r} charged-hadron data with NLO predictions
based on the MRST2 \protect\cite{4r} parton densities.
}
\label{fig14ps}
\end{center}
\end{figure}

\end{document}